\documentclass{article}
\usepackage{setspace}

\usepackage{doublespace}

\usepackage{graphics}

\begin{document}
\begin{spacing}{1.0}

\title{Energy mechanism of charges analyzed in real current environment}
\date{}
\maketitle

Reuven Ianconescu

26 Rothenstreich Str., Tel-Aviv, Israel

r\_iancon@excite.com\\

L .P. Horwitz

School of Physics and Astronomy, Raymond and Beverly Sackler,

Faculty of Exact Sciences, Tel Aviv University,

Ramat Aviv 69978, Israel

larry@post.tau.ac.il\\
\\
\\

\noindent We analyze in this work the energy transfer process of accelerated
charges, the mass fluctuations accompanying this process, and their
inertial properties. Based on a previous work, we use here the dipole
antenna, which is a very convenient framework for such analysis, for
analyzing those characteristics.
We show that the radiation process can be viewed by two energy
transfer processes: one from the energy source to the charges and the second
from the charges into the surrounding space. Those processes, not being in
phase, result in mass fluctuations. The same principle is true during
absorption. We show that in a transient period between absorption and
radiation the dipole antenna gains mass according to the amount of absorbed
energy and loses this mass as radiated energy. We rigorously prove that the
gain of mass, resulting from electrical interaction has inertial
properties in the sense of Newton's third low. We arrive to this
result by modeling the reacting spacetime region by an electric
dipole.\\


\\
\\

\noindent{\bf\large 1. INTRODUCTION}\\

\noindent
The common belief that an accelerating charge always radiates energy is
reviewed.
There are two Lienard-Wiechert potentials: one retarded and one
advanced. For reasons of causality, the advanced solution
is usually disregarded. It is considered as a field which ``knows'' the
future motion of a charged particle. But another way of interpreting the
advanced field is to define it as the field which {\it establishes} the future
motion of the particles.

Let us imagine a stationary charged particle, and some remote source of
radiation, distributed spatially in such a way as to create radiation
of the same pattern as a dipole antenna, but propagating inwards towards
the particle. May such a field correspond exactly to the advanced
field of {\it one} charged particle? The answer here is {\it no},
because the advanced field contains the Coulomb component, and there is
no way to create a Coulomb field from a far distributed source (because
of the Gauss law). In other words it is obvious that the advanced field
cannot create the state of a {\it single} charged particle.

But if we modify the above question: may such a field correspond
exactly to the advanced field of a neutral distribution of charged
particles? Such a distribution is Coulomb free, and therefore the answer
here is {\it yes}.  The best example of such a charge distribution is an
antenna. Also a single electric dipole is a good approximation to such
a charge distribution. And one may state that as long as the advanced
field does not interact with the above charge distribution, there is
{\it no} charge (because the interaction domain is neutral). The charge
becomes evident during the interaction. In this case one may state
that the advanced field {\it creates} the charges. Or, in other words
the advanced field converges into the future space-time domain of the
charges, and creates a new body defined by the charge distribution.

It is to be mentioned that considering the interaction region as totally
neutral before the interaction occurs, does not reduce the generality of the
problem, because we assume the universe is neutral, and hence every charge in
nature has its opposite sign counterpart. Therefore every space-time region is
neutral on a {\it sufficiently large scale}. In this work we consider the
interaction region to be large enough to validate the neutral system
assumption; alternatively, we may restrict ourselves to systems which are
locally neutral.

Where does this energy go now? If this charge distribution antenna is
connected to a resistor which equals the radiation resistance of
the antenna (i.e. impedance-matched), the energy transforms into heat.
If the antenna is in open circuit (or shorted), the energy is radiated back
into the space, on the same radiation pattern, and we will show that
during the period of interaction, the antenna gains inertial mass.

\begin{figure}[h]
{\par\centering \includegraphics{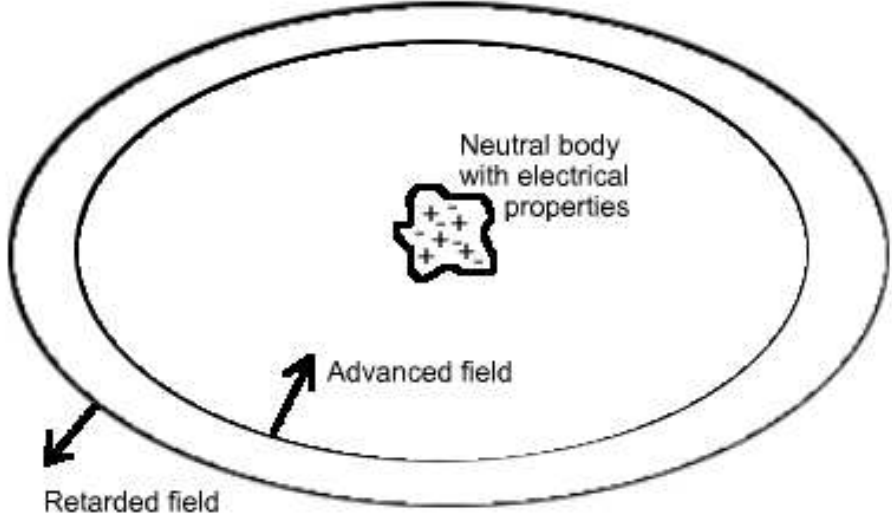} \par}
\caption{Advanced - incoming energy and retarded outgoing energy}
\label{generalconfig}
\end{figure}

This energy radiated back into the surrounding space will be
represented by the retarded field, i.e. the field {\it caused} by the
past charged distribution, or caused by the past (electric) existence
of the body, as shown in Figure \protect\ref{generalconfig}.


So interpreting the advanced field as an incoming energy (according
to the definition of the advanced Green's function), and
symmetrically the retarded field as an outgoing energy, and knowing
that both fields depend on acceleration of charges, we may generalize
the statement: ``an accelerating charge always radiates energy'', to:
accelerating {\it charges} always {\it radiate} or {\it absorb} energy.
There are two novel ideas here. One is that acceleration may indicate
radiation {\it or} absorption, i.e. interaction
\protect\cite{Feinmann-Wheeler}, and
the second is that the interaction is a collective process during which
charges accumulate energy from a source and release it to a destination,
and {\it not} a single charge process by which a charge ``radiates its
energy away''.

The known expression for the radiated power of a charged particle is
$P=2/3q^2a^{\mu}a_{\mu}$. For an on-shell particle,
$a^{\mu}v_{\mu}\equiv 0$, therefore we have the identity
$a^{\mu}a_{\mu}\equiv -\dot{a}^{\mu}v_{\mu}$. So the radiated power
could be expressed as $P=-2/3q^2\dot{a}^{\mu}v_{\mu}$. Interactions
always occur slightly off-shell \protect\cite{off_shell} (example:
if gravitation acts like acceleration, and under the equivalence principle
gravitation changes the metric, acceleration should change
the metric too), so during interactions the above
on-shell identity ($a^{\mu}a_{\mu}=-\dot{a}^{\mu}v_{\mu}$) is not
exactly satisfied.

One should therefore ask which one of the two expressions represents better
the radiated power $P$? It comes out that radiated power cannot come
from the charge itself, but from some source which boosted the charge (for
example a charge passing through a static electric field will accelerate,
by absorbing energy from the static field, and radiate it). We will show
that while $a^{\mu}a_{\mu}$ is responsible for the radiation,
$-\dot{a}^{\mu}v_{\mu}$ is responsible for the energy absorption {\it from}
the source, and therefore the difference between them,
$a^{\mu}a_{\mu}+\dot{a}^{\mu}v_{\mu}$, represents a transient mass
fluctuation.

In Section 2 we present the main results of a previous work
\protect\cite{prev_paper}, in which we analyzed the physical
meaning of the self force on a charge which is part of a radiating
antenna. This self force was associated with the radiation
resistance of the antenna. We will bring here also some new insights and
interpretations.

In Section 3 we develop the self force on a charge which is part of
a radiating or absorbing antenna. We will show here that while for
the radiating case, the self force generates the radiation resistance,
for the impedance-matched absorbing antenna the self force is responsible
for the dissipated heat.

In Section 4 we present an open circuit antenna which receives a short
energy pulse from the surrounding space and radiates it back. We show
that during the interaction process, the antenna gains an addition of
inertial mass which equals the received energy, and loses the
additional mass after radiation.

In Section 5, we explicitly show how the above mass gain has inertial
properties.\\

\noindent{\bf\large 2. SELF-FORCE ACTING AS RADIATION RESISTANCE}\\

\noindent
In a previous work \protect\cite{prev_paper} it has been shown that the
so called ``self force''
$F_{self}=(2/3)q^2\dot{a}$ is responsible for the radiation resistance
of an antenna. We will bring in this section, for convenience, the main
results of this work, plus some new interpretations.

The following problem, of a driven short dipole antenna of length $L$,
was formulated:

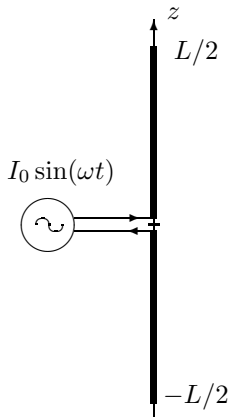
\begin{figure}[h]
\begin{picture}(170,170)(-50,0)
\linethickness{0.5 pt}
\put(125,0){\vector(0,1){150}}
\linethickness{2 pt}
\put(125,75){\line(0,1){65}}
\put(125,5){\line(0,1){65}}
\linethickness{0.5 pt}
\put(123,72.5){\line(1,0){4}}
\put(95,75){\line(1,0){31}}
\put(95,70){\line(1,0){31}}

\put(95,75){\vector(1,0){25}}
\put(126,70){\vector(-1,0){11}}

\put(85,72.5){\circle{18}}
\put(83,72.5){\oval(5,5)[t]}
\put(88,72.5){\oval(5,5)[b]}
\put(80,90){\makebox(20,5){$I_0\sin(\omega t)$}}
\put(130,150){\makebox(5,5){$z$}}
\put(132,135){\makebox(18,5){$L/2$}}
\put(132,5){\makebox(18,5){$-L/2$}}
\end{picture}
\caption{Short dipole antenna}
\label{dipoleantenna}
\end{figure}

Given the fact that any time dependence can be expanded in a Fourier series
of functions, we consider the harmonic time dependence, as shown in
Figure \protect\ref{dipoleantenna}. But the results we obtained for the fields
are correct in general, and we use the harmonic dependence only when
calculating the radiation resistance, which is indeed frequency dependent.

The $z$ dependence of the current is disregarded, the antenna being short.

It comes out therefore that the motion parameters (velocity, acceleration,
and its derivative) of the charges have the form

\begin{equation}
v=v_0\sin(\omega t); a=\omega v_0\cos(\omega t); \dot{a}=-\omega^2 v_0\sin(\omega t)
\label{timedep}
\end{equation}

\begin{figure}[h]
\begin{picture}(150,150)(-50,0)
\linethickness{0.5 pt}
\put(100,135){\vector(0,1){15}}
\put(105,150){\makebox(8,5){$z$}}

\linethickness{2 pt}
\put(80,0){\line(0,1){135}}
\put(120,0){\line(0,1){135}}
\put(100,90){\circle*{8}}
\put(100,65){\circle*{8}}
\put(100,40){\circle*{8}}
\put(100,15){\circle*{8}}

\linethickness{0.5 pt}
\put(40,140){\line(1,-1){15}}
\put(55,125){\line(1,1){10}}
\put(65,135){\vector(1,-1){13}}

\put(140,140){\line(-1,-1){35}}
\put(105,105){\line(2,-1){10}}
\put(115,100){\vector(-4,-3){10}}

\put(40,65){\vector(1,0){55}}
\put(40,40){\vector(1,0){55}}

\put(40,70){\makebox(25,5){Charge B}}
\put(40,45){\makebox(25,5){Charge A}}

\put(125,65){\line(1,0){25}}
\put(125,40){\line(1,0){25}}

\put(132,60){\vector(0,1){5}}
\put(132,45){\vector(0,-1){5}}

\put(134,50){\makebox(30,5){$\Delta z=q/\rho$}}

\put(40,145){\makebox(25,5){conductor}}
\put(143,145){\makebox(25,5){free charges}}
\put(147,137){\makebox(25,5){of size $q$}}

\end{picture}
\caption{Close-up view of the conductor}
\label{discretization}
\end{figure}
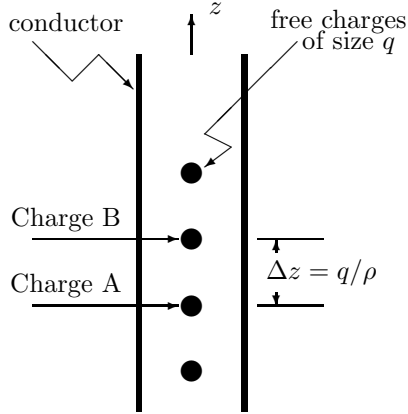

By describing the conductor as a continuum of single charges, as in
Figure \protect\ref{discretization}, we examined the forces on the ``test charge'' B, as a result of
a disturbance produced on charge A. This disturbance is expressed in terms
of the motion parameters of charge A, i.e. the acceleration and its
derivative.

The motion parameters are connected to the current via:

\begin{equation}
I=v\rho=qv/\Delta z
\label{micromacro}
\end{equation}

where $I$ is the electrical current and $\rho$ is the free charge density
of the conductor. The discretization of the continuum into single charges
is done by defining discrete charges of value $q$ at distance $\Delta z$
so that $q=\rho\Delta z$, as in Figure \protect\ref{discretization}.
Furthermore, the velocity of
the charges in a conductor is extremely small (order of magnitude of
$10^{-4} m/sec$).

The result obtained for the disturbing field was
(\protect\ref{retarded_field}):

\begin{equation}
E_z = q\left(\frac{1}{\Delta z^2}-  \frac{a}{\Delta z} + \frac{2}{3} \dot{a}\right)
\label{retarded_field}
\end{equation}

It was shown \protect\cite{prev_paper} that the $\dot{a}$ term in
(\protect\ref{retarded_field}), being in opposite phase with the
velocity of the charge, and the velocity being in phase with the
current, is the {\it only} term in (\protect\ref{retarded_field})
representing the damping force for the source.

The damping force is

\begin{equation}
F_{self}=(2/3)q^2\dot{a}
\label{damping_force}
\end{equation}

therefore the ``self'' work (or power) is

\begin{equation}
P_{self}=-F_{self}v=-(2/3)q^2\dot{a}v
\label{damping_power}
\end{equation}

and the radiation resistance, defined as $\frac{P_{self}}{I^2}$, using
(\protect\ref{timedep}) and (\protect\ref{micromacro}), comes
out:

\begin{equation}
R=(2/3)(2\pi)^2(\Delta z/\lambda)^2
\label{rad_resistance}
\end{equation}

The magnitude $\Delta z=L$ represents the short dipole antenna length, 
and the wavelength $\lambda=2\pi/\omega$.

Another way to look at $R$ is to define the ``self'' voltage as an integral
on the ``self'' field $(2/3)q^2\dot{a}$:

\begin{equation}
V_{self}=-\int E_{self}dz \simeq -E_{self}\Delta z=-(2/3)q^2\dot{a}\Delta z
\label{self_voltage}
\end{equation}

and obtain $R$ as $V_{self}/I$.

It is remarkable that the ``self force'', calculated with the {\it micro}
parameters, shows up in {\it macro} to have the meaning of a damping power
which is responsible for the radiation resistance.

Up to here, we have given the main results of \protect\cite{prev_paper}.
Let us now compare between the ``self power'', which was shown to act
as damping power, and the radiated power.

The power radiated by a (low velocity) charge is

\begin{equation}
P_{rad}=(2/3)q^2a^2
\label{rad_power}
\end{equation}

The ``self power'' equals {\it identically} $I^2 R$, so this is the
power the current source pumps. Therefore, it
represents the power generated by the current source, and pumped into
the charges. On the other hand, the radiated power represents
the power which the charges release into space.

Using (\protect\ref{timedep}), we see that
$P_{self}=(2/3)q^2(\omega v_0)^2\cos^2(\omega t)$ and
$P_{rad}=(2/3)q^2(\omega v_0)^2\sin^2(\omega t)$. We remark that
the energy over an entire cycle $2\pi/\omega$ of both powers is identical,
so the energy is conserved {\it in average}, however there is a phase shift
between them, suggesting that the charges accumulate energy, before
radiating it into space.

Using (\protect\ref{damping_power}) and (\protect\ref{rad_power}), the power
accumulated by the charges is

\begin{equation}
W=P_{self}-P_{rad}=-(2/3)q^2(a^2+\dot{a}v)=-(2/3)q^2(v_0\omega)^2\cos(2\omega t)\equiv -P_0\cos(2\omega t)
\label{acc_power}
\end{equation}

where $P_0$ is easily shown to be equal to $RI_0^2$ and is also the peak
value of $P_{self}$ or $P_{rad}$ (which are equal up to a phase).

It is generally assumed that mass shell remains fixed classically, and varies
only in quantum mechanics; we disagree because the on shell condition
$v^{\mu}v_{\mu}=-1$ always imposes constraints which might contradict
reality. For example it imposes the constraint $F^{\mu}v_{\mu}=0$
for any force $F^{\mu}=ma^{\mu}$ in Newton's law.

Therefore, interacting particles should not be completely on-shell
\protect\cite{off_shell}, so we will interpret (\protect\ref{acc_power})
by considering the charges slightly off-shell:

\begin{equation}
v^{\mu}v_{\mu}=-1+v^2
\label{v_squared}
\end{equation}

The reason for choosing the off-shell factor $v^2$ will become clear
later, and it is to be mentioned that $v^2\ll 1$, as we stated
already. Taking the derivative of the left side with respect to the
proper time, and the derivative of the right side with respect to the
time (supposing time and proper time are almost equivalent), we
obtain:

\begin{equation}
a^{\mu}v_{\mu}=va
\label{va}
\end{equation}

where $a=\dot{v}$. By taking again the derivative, we obtain:

\begin{equation}
\dot{a}^{\mu}v_{\mu}+a^{\lambda}a_{\lambda}=a^2+\dot{a}v
\label{av_plus_a_squared}
\end{equation}

The relativistic expression of the self force
$F_{self}^{\mu}=(2/3)q^2(\dot{a}^{\mu}-a^{\lambda}a_{\lambda}v^{\mu}$),
multiplied by the particle's velocity $v_{\mu}$, results in identically $0$
on-shell, but in our case it gives:

\begin{equation}
F_{self}^{\mu}v_{\mu}=(2/3)q^2(\dot{a}^{\mu}v_{\mu}-a^{\lambda}a_{\lambda}v^{\mu}v_{\mu})\simeq (2/3)q^2(a^2+\dot{a}v)
\label{covariant_power}
\end{equation}

The right side of (\protect\ref{covariant_power}) has been obtained using
(\protect\ref{av_plus_a_squared}), and the fact that $v\ll 1$.

We remark that $F_{self}^{\mu}v_{\mu}=-W$ in (\protect\ref{acc_power}).
The mass of a radiating antenna is therefore:

\begin{equation}
m(t)=-\int F_{self}^{\mu}v_{\mu}d\tau \simeq -(2/3)q^2\int(a^2+\dot{a}v)dt=-(2/3)q^2av+M=-\frac{P_{eff}}{\omega}\sin(2\omega t)+M
\label{mass_harmonic}
\end{equation}

where $M$ is the integration constant and represents the interaction-free
mass of the antenna, and $P_{eff}=P_0/2$ represents the average radiated power.

It is obvious that the free mass is preserved in average, because we dealt
here with a stationary case of harmonic signal for which energy is pumped
into the charges and almost immediately radiated into the space.

The mass variation will be further examined in Section 4, for a short
excitation pulse.\\

\noindent{\bf\large 3. ANTENNA ABSORBING ADVANCED FIELDS AND RADIATING RETARDED FIELDS}\\

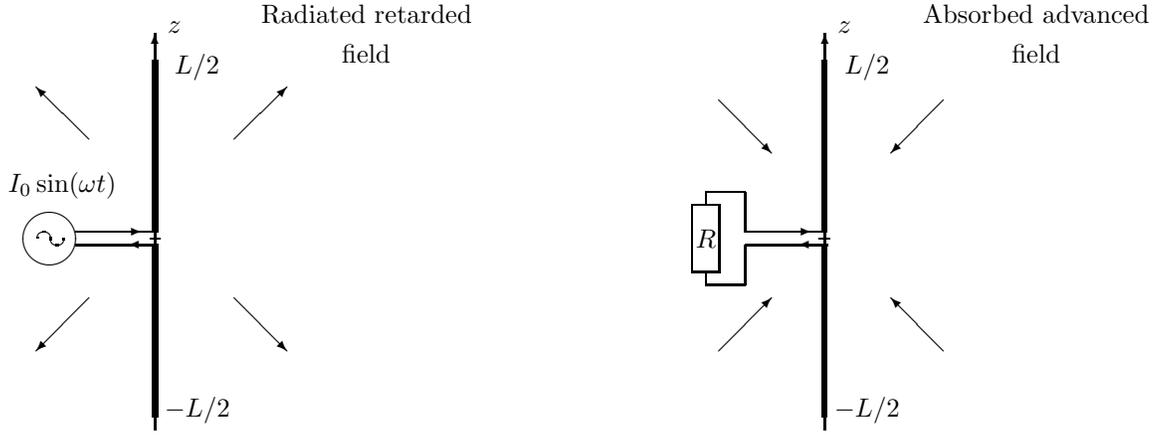
\begin{figure}[h]
\begin{picture}(170,170)(80,0)
\linethickness{0.5 pt}
\put(125,0){\vector(0,1){150}}
\linethickness{2 pt}
\put(125,75){\line(0,1){65}}
\put(125,5){\line(0,1){65}}
\linethickness{0.5 pt}
\put(123,72.5){\line(1,0){4}}
\put(95,75){\line(1,0){31}}
\put(95,70){\line(1,0){31}}

\put(95,75){\vector(1,0){25}}
\put(126,70){\vector(-1,0){11}}

\put(85,72.5){\circle{18}}
\put(83,72.5){\oval(5,5)[t]}
\put(88,72.5){\oval(5,5)[b]}
\put(80,90){\makebox(20,5){$I_0\sin(\omega t)$}}
\put(130,150){\makebox(5,5){$z$}}
\put(132,135){\makebox(18,5){$L/2$}}
\put(132,5){\makebox(18,5){$-L/2$}}

\put(155,110){\vector(1,1){20}}
\put(155,50){\vector(1,-1){20}}
\put(100,110){\vector(-1,1){20}}
\put(100,50){\vector(-1,-1){20}}
\put(195,155){\makebox(20,5){Radiated retarded}}
\put(195,140){\makebox(20,5){field}}

\end{picture}
\begin{picture}(170,170)(0,0)
\linethickness{0.5 pt}
\put(125,0){\vector(0,1){150}}
\linethickness{2 pt}
\put(125,75){\line(0,1){65}}
\put(125,5){\line(0,1){65}}
\linethickness{0.5 pt}
\put(123,72.5){\line(1,0){4}}
\put(95,75){\line(1,0){31}}
\put(95,70){\line(1,0){31}}

\put(95,75){\vector(1,0){25}}
\put(126,70){\vector(-1,0){11}}

\put(95,75){\line(0,1){15}}
\put(95,70){\line(0,-1){15}}

\put(95,90){\line(-1,0){15}}
\put(95,55){\line(-1,0){15}}

\put(80,90){\line(0,-1){5}}
\put(80,55){\line(0,1){5}}

\put(75,60){\framebox(10,25){$R$}}

\put(130,150){\makebox(5,5){$z$}}
\put(132,135){\makebox(18,5){$L/2$}}
\put(132,5){\makebox(18,5){$-L/2$}}

\put(170,125){\vector(-1,-1){20}}
\put(170,30){\vector(-1,1){20}}
\put(85,125){\vector(1,-1){20}}
\put(85,30){\vector(1,1){20}}
\put(195,155){\makebox(20,5){Absorbed advanced}}
\put(195,140){\makebox(20,5){field}}

\end{picture}
\caption{Radiating or Absorbing dipole antenna}
\label{antenna_adv_ret}
\end{figure}

\noindent
The formulation here is quite similar to this in \protect\cite{prev_paper},
and it is shown in Figure \protect\ref{antenna_adv_ret}.

As we saw, the ``self'' retarded field of a charge, acting on its neighbor,
generates for a radiating antenna a ``self'' voltage which, divided by
the current results in the radiation resistance.

It is therefore natural to analyze the ``self'' advanced field of a charge,
acting on its neighbor, for an absorbing antenna.

We therefore derive the the retarded and advance field on charge B, by the
motion of the disturbed charge A.

The near field of a charge is expressed by \protect\cite{prev_paper}

\begin{equation}
E_z = \frac{q}{[R_\mu v^\mu]^2}_{\tt RET/ADV}
\label{near_field}
\end{equation}

where $R_\mu$ is the null vector from charge A to charge B, given by

\begin{equation}
R_\mu=(-t_A(\tau),0,0,z_B-z_A(\tau))
\label{R_mu}
\end{equation}

The null vector condition is imposed by

\begin{equation}
t_A^2=|z_B-z_A(t_A)|^2
\label{null_vector_condition0}
\end{equation}

hence

\begin{equation}
{t_A}_{\tt RET/ADV}= \mp(z_B-z_A(t_A))
\label{null_vector_condition}
\end{equation}

Expressing $v_\mu=\gamma(1,0,0,v_A(\tau))$, knowing that $\gamma\simeq 1$,
we may express $R_\mu v^\mu$ as

\begin{equation}
R_\mu v^\mu=\mp(z_B-z_A)(1\mp v_A)
\label{R_mu_v_mu}
\end{equation}

The position $z_A$ of charge A may be expanded in series of $t_A$

\begin{equation}
z_A(t_A)=z_{A_0}+\frac{1}{2}at_A^2+\frac{1}{6}\dot{a}t_A^3
\label{z_A}
\end{equation}

where $z_{A_0}$, $a$ and $\dot{a}$ are the position, acceleration and its
derivative of charge A at $t=0$, respectively, and we chose the reference
frame so that the velocity is $0$ at $t=0$.

We may calculate now $t_A$ which satisfies the null vector condition.
By putting (\protect\ref{z_A}) in (\protect\ref{null_vector_condition}),
we obtain

\begin{equation}
z_B-z_A(t_A)=\Delta z_0-\frac{1}{2}at_A^2-\frac{1}{6}\dot{a}t_A^3=\mp t_A
\label{null_vector_condition1}
\end{equation}

where $\Delta z_0\equiv z_B-z_{A_0}$. The the null vector condition may be
rewritten as follows

\begin{equation}
\Delta z_0\pm t_A\left(1 \mp \frac{1}{2}at_A \mp \frac{1}{6}\dot{a}t_A^2\right)=0
\label{null_vector_condition2}
\end{equation}

This may be solved in first approximation by
${t_A}_{\tt RET/ADV}=\mp\Delta z_0$.
By setting the first approximation solution into
(\protect\ref{null_vector_condition2}), we obtain the retarded/advanced
solution for $t_A$

\begin{equation}
{t_A}_{\tt RET/ADV}=\frac{\mp \Delta z_0}{1 + \frac{1}{2}a\Delta z_0 \mp \frac{1}{6}\dot{a}\Delta z_0^2}
\label{t_A}
\end{equation}

Using (\protect\ref{null_vector_condition1}) in (\protect\ref{R_mu_v_mu})
we obtain

\begin{equation}
R_\mu v^\mu=t_A(1\mp v_A)
\label{R_mu_v_mu1}
\end{equation}

We may expand $v_A$:

\begin{equation}
v_A=at_A+\frac{1}{2}\dot{a}t_A^2
\label{v_a}
\end{equation}

and obtain

\begin{equation}
R_\mu v^\mu=t_A \mp at_A^2 \mp \frac{1}{2}\dot{a}t_A^3
\label{R_mu_v_mu11}
\end{equation}

Using (\protect\ref{t_A}), we expand $t_A$, $t_A^2$ and $t_A^3$ in powers of
$\Delta z_0$, keeping terms up to $\Delta z_0^3$:

\begin{equation}
{t_A}=\mp \Delta z_0 \left(1 - \frac{1}{2}a\Delta z_0 \pm \frac{1}{6}\dot{a}\Delta z_0^2 + \frac{1}{4} a^2 \Delta z_0^2\right)
\label{t_A1}
\end{equation}

\begin{equation}
{t_A^2}=\Delta z_0^2 (1 - a\Delta z_0)
\label{t_A2}
\end{equation}

\begin{equation}
{t_A^3}=\mp \Delta z_0^3
\label{t_A3}
\end{equation}

Setting those values into (\protect\ref{R_mu_v_mu11}), we obtain

\begin{equation}
R_\mu v^\mu=\mp\Delta z\left(1+\frac{1}{2}a\Delta z \mp \frac{1}{3}\dot{a}\Delta z^2-\frac{3}{4}a^2\Delta z^2\right)
\label{R_mu_v_mu2}
\end{equation}

here we replace $\Delta z_0$ by $\Delta z$, which is the average distance
between charges, as described in Figure \protect\ref{discretization}.

Putting now (\protect\ref{R_mu_v_mu2}) into (\protect\ref{near_field}), we
obtain

\begin{equation}
E_z = q (\frac{1}{\Delta z^2}-\frac{a}{\Delta z} \pm \frac{2}{3} \dot{a} + \frac{9}{4}a^2)
\label{E_z}
\end{equation}

Dealing with harmonic excitation, $\dot{a}=\ddot{v} \sim \omega^2 v$, and
$(\dot{v})^2 \sim (\omega v)^2$. Having $v$ smaller by 11 orders of magnitude
than light velocity, $\dot{v}^2$ is completely negligible relative to
$\ddot{v}$, for any frequency,  and we obtain:

\begin{equation}
E_z = q \left(\frac{1}{\Delta z^2}-\frac{a}{\Delta z} \pm \frac{2}{3} \dot{a}\right)
\label{ret_adv_field}
\end{equation}

where the upper sign refers to retarded and the lower sign refers to
advanced (compare with (\protect\ref{retarded_field})).

The first term $q/\Delta t^2$ {\it is always cancelled} by the
force of the ``other'' neighbor, it therefore can be completely {\it ignored}.

The third term is identical to what is considered to be the field which
creates the retarded/advanced self-force of a charge, but here it was derived
as the force on a charge, due to a disturbance on a neighboring charge.
As we shall see, it is the {\it only} term which responsible for the
radiation/absorption resistance (which is a local phenomenon on the world
line).

We will call the last 2 terms of (\protect\ref{ret_adv_field}) $E_{self}$,
obtaining

\begin{equation}
E_{self} = q \left(-{\frac{a}{\Delta z}} \pm \frac{2}{3} \dot{a}\right)
\label{ret_adv_self_field}
\end{equation}

\noindent The potential difference $V$ on a wire segment of length $\Delta z$,
resulting from $E_{self}$, will be denoted as the ``self'' tension (or ``self''
voltage) and is calculated by

\begin{equation}
V=-E_{self}\Delta z
\label{ret_adv_self_tension}
\end{equation}

\noindent The magnitude $\Delta z=q/\rho$ is according to the discretization
of the charge distribution, so we get from (\protect\ref{ret_adv_self_field}) and
(\protect\ref{ret_adv_self_tension})

\begin{equation}
V = \mp {\frac{q^2 [2/3 \ddot{v}]}{\rho}}  + q \dot{v}
\label{ret_adv_self_tension1}
\end{equation}

\noindent According to (\protect\ref{micromacro})

\begin{equation}
\dot{v}=\frac{1}{\rho} \frac{\partial I}{\partial t} \ and \ \ddot{v}=\frac{1}{\rho}\frac{\partial^2 I}{\partial t^2}
\label{v_and_ddot_v_func_I}
\end{equation}

\noindent Dealing with harmonic excitation (see (\protect\ref{timedep}))
$\partial / \partial t$ is like multiplication by $\omega$ (up to a $90^0$
phase).

We therefore may write (\protect\ref{ret_adv_self_tension1}), using
(\protect\ref{v_and_ddot_v_func_I})

\begin{equation}
V = \mp \frac{2}{3} \left(\frac{q}{\rho}\right)^2 \frac{\partial^2 I}{\partial t^2} + \frac{q}{\rho} \frac{\partial I}{\partial t} = \mp \frac{2}{3} \Delta z^2 \frac{\partial^2 I}{\partial t^2} + \Delta z \frac{\partial I}{\partial t}
\label{ret_adv_self_tension2}
\end{equation}

\noindent The power radiated/absorbed by the segment $\Delta z$ is
$\Delta P=VI$ and given by

\begin{equation}
\Delta P= \mp \frac{2}{3} \frac{\partial^2 I}{\partial t^2} I \Delta z^2 + \frac{\partial I}{\partial t} I \Delta z
\label{delta_p}
\end{equation}

According to (\protect\ref{timedep}),
$\frac{\partial^2 I}{\partial t^2} = -\omega^2 I$.
The constant ratio between
$\partial^2 I / \partial t^2$ and $I$ means that the current is ``in
phase'' with its second derivative, and therefore the
first part of $\Delta P$ in (\protect\ref{delta_p}),
integrated over time, represents radiated or absorbed energy.
However, the multiplication of $I$ by $\partial I / \partial t$
has the form of $\cos\omega(t)\sin\omega(t)$ and
therefore the second part of $\Delta P$ represents a {\it reactive}
power, which results in zero energy after integrating on an integer
number of time cycles. The reactive power represents power which is
returned to the source each time cycle. We are therefore interested in
the first term of $\Delta P$ in (\protect\ref{delta_p}).

We therefore obtain

\begin{equation}
P_{self-ret/adv}= \pm \frac{2}{3} \omega^2 (I \Delta z)^2=\mp (2/3)q^2\dot{a}v
\label{damping_excite_power}
\end{equation}

(compare with (\protect\ref{damping_power})). The radiation/absorption
resistance is obtained by $P_{self-ret/adv}/I^2$ or by the first part
of $V$ in (\protect\ref{ret_adv_self_tension}) divided by the current $I$:

\begin{equation}
R_{rad/abs}=\pm (2/3)(2\pi)^2(\Delta z/\lambda)^2
\label{rad_abs_resistance}
\end{equation}

The radiation resistance is the same as in (\protect\ref{rad_resistance}),
and the absorption resistance is obviously its negative counterpart.

This is because a negative resistance $-R$ through which a current $I$
flows is {\it completely equivalent} to a current source $I$ with an internal
resistance $R$. If we connect a load resistance of the same value $R$
to this current source, as in the right side of
Figure \protect\ref{antenna_adv_ret}, we get an impedance-matched
{\it absorbing} antenna.

We proved here that the advanced field behaves like an absorbed field,
and knowing that the Poynting vector associated with the advanced field
is of the same size and opposite direction as the Poynting vector
associated with the retarded field, we may clearly state that the absorbed
power is

\begin{equation}
P_{abs}=-(2/3)q^2a^2
\label{abs_power}
\end{equation}

(compare with (\protect\ref{rad_power})).

From here on we may follow the same arguments from Section 2, from
(\protect\ref{rad_power}) to (\protect\ref{mass_harmonic}), and get
the analogous results for an absorbing antenna, which are basically
the results for a radiating antenna, with sign changed.

The power accumulated by the charges is

\begin{equation}
W=P_{self-adv}-P_{abs}=(2/3)q^2(a^2+\dot{a}v)=(2/3)q^2(v_0\omega)^2\cos(2\omega t)\equiv P_0\cos(2\omega t)
\label{acc_power_abs_antenna}
\end{equation}

where $P_0$ is the peak value of $P_{self}$ or $P_{abs}$
(which are equal up to a phase).

The mass of an absorbing antenna comes out to be

\begin{equation}
m(t)=-\int F_{self}^{\mu}v_{\mu}d\tau \simeq (2/3)q^2\int(a^2+\dot{a}v)dt=(2/3)q^2av+M=\frac{P_{eff}}{\omega}\sin(2\omega t)+M
\label{mass_harmonic_abs_antenna}
\end{equation}

where $M$ is the interaction-free mass of the antenna, and
$P_{eff}=P_0/2$ represents the average absorbed power.

We see that the mass of an antenna is expressed via the integral on
the self force of its off-shell charges
$m(t)=-\int F_{self}^{\mu}v_{\mu}d\tau$, and the difference between the
radiating and absorbing cases is the sign. This sign difference is
associated with the fact that $F_{self-adv}^{\mu}=-F_{self-ret}^{\mu}$.\\

\noindent{\bf\large 4. THE TRANSIENT FROM ABSORPTION TO RADIATION}\\

\noindent
We saw in the Sections 2 and 3 that the mass of an antenna radiating or
absorbing an harmonic signal, oscillates harmonically at twice the
radiating/absorbing frequency.

The harmonic signal has no local time properties, hence the above
results cannot describe transients between emission and absorption.

We analyze here the case of an antenna of whose discretized charges
have a constant acceleration $a_z$ for a short period of time $2\Delta t$.
As we saw in Section 3, acceleration can indicate radiation {\it or}
absorption, so we will consider the case for which in the first half of the
acceleration duration (time from $-\Delta t$ to $0$) there is absorption,
and in the second half of the acceleration duration (time from
$0$ to $\Delta t$) there is radiation.

The retarded/advance field due to a discretized charge in Figure
\protect\ref{discretization} at the location of its neighbor is
given by the last 2 terms in (\protect\ref{ret_adv_field}):

\begin{equation}
E_z = q \left(-\frac{a}{\Delta z} \pm \frac{2}{3} \dot{a}\right)
\label{ret_adv_field1}
\end{equation}

where the upper sign refers to retarded and the lower sign refers to
advanced, and the first term in (\protect\ref{ret_adv_field}) $q/\Delta t^2$
{\it is always cancelled} by the force of the ``other'' neighbor.

The force on the neighbor charge $q_1$ will therefore be

\begin{equation}
F_z = q_1 E_z = q_1 q \left(-\frac{a}{\Delta z} \pm \frac{2}{3} \dot{a}\right) = \frac{-q_1 q}{\Delta z} a \left(1 \mp (2/3) \frac{\dot{a}}{a}\Delta z\right) \equiv Ma\left(1 \mp (2/3) \frac{\dot{a}}{a}\Delta z\right)
\label{ret_adv_force}
\end{equation}

where the lower sign refers to the absorption time interval
($-\Delta t$ to $0$), and the upper sign to the radiation time
interval ($0$ to $\Delta t$).

Here we defined the mass $M$ as minus the product of two neighboring charges
over the distance between them. We shall see in Section 5 that such a
configuration of a pair of charges has inertial properties of magnitude
$-q_1 q/{\Delta z}$, hence when the charges are opposite in sign they
behave like a positive mass, and {\it vice versa}.

Actually our model in Figure \protect\ref{discretization} considered
equal (free) charges in a conductor, but we know that for each negative
charge there is a positive counterpart, so we will consider for now
$M$ in (\protect\ref{ret_adv_force}) as positive: $M\equiv q^2/\Delta z$.

According to (\protect\ref{ret_adv_force}), the dynamic mass is

\begin{equation}
m(t) = M\left(1 \mp (2/3) \frac{\dot{a}(t)}{a(t)}\Delta z\right)
\label{dynamic_mass}
\end{equation}

We want to investigate the behavior of the dynamic mass change during the
interaction, i.e. from time $-\Delta t$ to $\Delta t$. For
$-\Delta t<t<0$, an absorption occurs, therefore the lower sign has to
be used, and for $0<t<\Delta t$, the upper sign has to be used.

A fixed acceleration $a_z$ means that if we go to the {\it rest} frame of the
particle by Lorentz transformation at any proper time, and measure the
acceleration in this frame,
we always get the same result $a_z$. Under this definition, $\dot{a}$ is not
identically $0$. A fixed accelerated motion, in the $z$ direction may
be parametrized in the following way:

\begin{eqnarray}
\label{fixed_acceleration_motion_v}
v^{\mu}(\tau)=(\cosh(a_z\tau), 0, 0, \sinh(a_z\tau)) \\
\label{fixed_acceleration_motion_a}
a^{\mu}(\tau)=a_z(\sinh(a_z\tau), 0, 0, \cosh(a_z\tau)) \\
\label{fixed_acceleration_motion_dot_a}
\dot{a}^{\mu}(\tau)=a_z^2(\cosh(a_z\tau), 0, 0, \sinh(a_z\tau))
\end{eqnarray}

where the components of the 4-vectors are $(t,x,y,z)$.

This low order approximation maintains the mass shell constraint
\protect\cite{off_shell} $v^\mu v_\mu=-1$, neglecting the $v^2$
correction noted in (\protect\ref{v_squared}). But we shall see
below, that this off-shell term will {\it reappear}.

In the framework of a conductor, we showed \protect\cite{prev_paper} that the
velocity of the charges is of order of magnitude of $10^{-4} m/sec$, hence
$v\ll 1$, so we have to use the limit of
(\protect\ref{fixed_acceleration_motion_v})-(\protect\ref{fixed_acceleration_motion_dot_a})
around the apex, up to first order in $\tau$ (note that for $v\ll 1$,
$t\simeq\tau$)

\begin{eqnarray}
\label{fixed_acceleration_approx_v}
v^{\mu}(\tau)\simeq (1 , 0, 0, a_z t) \\
\label{fixed_acceleration_approx_a}
a^{\mu}(\tau)\simeq a_z(a_z t , 0, 0, 1) \\
\label{fixed_acceleration_approx_dot_a}
\dot{a}^{\mu}(\tau)\simeq a_z^2(1 , 0, 0, a_z t)
\end{eqnarray}

We remark that
(\protect\ref{fixed_acceleration_approx_v})-(\protect\ref{fixed_acceleration_approx_dot_a})
satisfy the off-shell condition
$v^\mu v_\mu=-1+v^2$ defined in (\protect\ref{v_squared}).

We divide now the $z$ component of $\dot{a}^{\mu}$, by the $z$ component
of $a^{\mu}$, to set it in (\protect\ref{dynamic_mass}), and get for the 
absorption period $-\Delta t<t<0$

\begin{equation}
m(t) = M(1 + (2/3) a_z^2 t\Delta z); \hspace{5pt} -\Delta t<t<0
\label{dynamic_mass_abs}
\end{equation}

and for the radiation period $0<t<\Delta t$

\begin{equation}
m(t) = M(1 - (2/3) a_z^2 t\Delta z); \hspace{5pt} 0<t<\Delta t
\label{dynamic_mass_rad}
\end{equation}

During the absorption period, there is a mass gain of
$(2/3)M a_z^2 t\Delta z$, and using $M=q^2/\Delta z$, the mass gain
is

\begin{equation}
m(0)-m(-\Delta t) = (2/3) q^2 a_z^2 \Delta t \equiv P_{absorbed}\Delta t
\label{mass_gain}
\end{equation}

and similarly, during the radiation period there is a mass {\it loss},
resulting in a negative mass gain:

\begin{equation}
m(\Delta t) - m(0) = - (2/3) q^2 a_z^2 \Delta t\equiv -P_{radiated}\Delta t
\label{mass_loss}
\end{equation}

We recognize in the expressions for the gain and lost of mass, the
absorbed and radiated energies, respectively, exhibiting energy
conservation.\\

\noindent{\bf\large 5. INERTIAL BEHAVIOR OF CHARGES}\\

\noindent
The analysis, done up to this point, was based on the discretization of
an antenna into equally spaced charges, attributing to a given charge
micro parameters ($v$, $a$, $\dot{a}$), connected to the antenna macro
parameters, like $I$ and $V$. We eventually considered for a short
antenna, the distance between 2 charges $\Delta z$, as a representative
length of the antenna.

With the aid of this model, we have succeeded to prove that
the retarded field behaves like a radiated field and the advanced field
behaves like an absorbed field. We have shown that during radiation or
absorption the mass of the antenna (defined as $M\equiv q^2/\Delta z$),
oscillates at twice the frequency.

When the antenna absorbs an energy pulse, the mass increases according
to the absorbed energy, and during the re-radiation of the pulse, the
mass decreases.

In the current section we wish to make a rigorous analysis to show that
an electrically reacting spacetime region has inertial properties. We
model this reacting spacetime region by an ideal electric dipole. We
define here an ideal dipole, as two equal and opposite charges at a
distance $d$, so that for any possible acceleration $a$ of this system:

\begin{equation}
ad \rightarrow 0
\label{ideal_dipole_condition}
\end{equation}

Note that $ad$ is unitless, therefore condition
(\protect\ref{ideal_dipole_condition}) is well defined. It means that
the distance between the charges is short enough for the largest
possible acceleration that may be considered.

\begin{figure}[h]
\begin{picture}(150,150)(-50,0)
\linethickness{0.5 pt}
\put(80,40){\vector(1,0){170}}
\put(250,35){\makebox(8,5){$z$}}

\put(80,40){\vector(0,1){90}}
\put(72,130){\makebox(8,5){$x$}}

\put(80,40){\vector(-1,-1){40}}
\put(33,-8){\makebox(8,5){$y$}}

\linethickness{2 pt}
\put(100,65){\circle*{8}}
\put(100,40){\circle*{8}}

\linethickness{0.5 pt}

\put(100,72){\makebox(8,5){$q_2$}}
\put(100,28){\makebox(8,5){$q_1$}}

\put(105,65){\line(1,0){25}}
\put(105,40){\line(1,0){25}}

\put(112,57){\vector(0,1){8}}
\put(112,47){\vector(0,-1){7}}

\put(110,50){\makebox(8,5){$d$}}

\put(140,52){\vector(1,0){35}}
\put(178,52){\makebox(8,5){$a_z$}}

\end{picture}
\caption{Accelerating dipole}
\label{accelerating_dipole}
\end{figure}
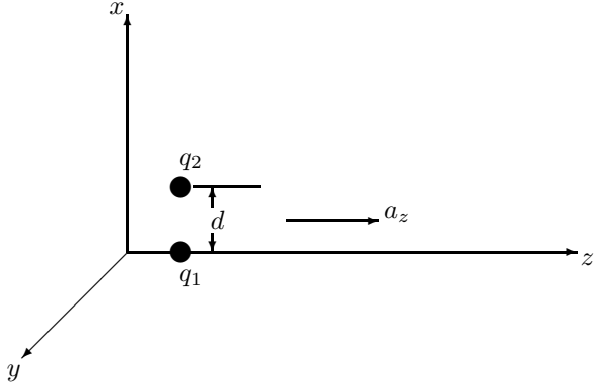

The formulation is based on Figure \protect\ref{accelerating_dipole}

In Figure \protect\ref{accelerating_dipole} the charges appear as $q_1$
$q_2$ but we will be eventually interested in the case of $q=q_1=-q_2$.
The charges are bound, and they move together on the $z$ axis.

We calculate the field on $q_2$, caused by $q_1$, in an inertial system
in which $q_2$ is at rest.

The most general expression for the electromagnetic tensor of a moving
charge $q$ is given by \protect\cite{Rohrlich}

\begin{equation}
F^{\mu\nu}=\pm\frac{q}{R^{\sigma}v_{\sigma}}\frac{d}{d\tau}\left(\frac{R^{\mu}v^{\nu}-R^{\nu}v^{\mu}}{R^{\sigma}v_{\sigma}}\right)
\label{em_tensor_exact}
\end{equation}

where $\pm$ refer to retarded and advanced respectively,
$R^{\mu}(\tau)\equiv X^{\mu}-Z^{\mu}(\tau)$, $X^{\mu}$ being the observation
point and $Z^{\mu}(\tau)$ the particle's world line. Expression
(\protect\ref{em_tensor_exact}) is calculated at $\tau$ which satisfies

\begin{equation}
R^{\sigma}R_{\sigma}=0
\label{ret_adv_condition}
\end{equation}

Equation (\protect\ref{ret_adv_condition}) has always two solutions for
$\tau$, the earlier corresponds to retarded and the later to advanced.

The values we shall set in (\protect\ref{em_tensor_exact}) for a constant
acceleration $a_z$ along the $z$ axis are

\begin{equation}
Z^{\mu}=a_z^{-1}(\sinh(a_z\tau), 0, 0, \cosh(a_z\tau))
\label{Z_mu}
\end{equation}

\begin{equation}
X^{\mu}=(0, d, 0, a_z^{-1})
\label{X_mu}
\end{equation}

because the observation point is defined as the location of $q_2$ when
it is at rest, i.e. at the apex of the hyperbola.

\begin{equation}
v^{\mu}=(\cosh(a_z\tau), 0, 0, \sinh(a_z\tau))
\label{v_mu}
\end{equation}

therefore

\begin{equation}
R^{\mu}=a_z^{-1}(-\sinh(a_z\tau), d, 0, 1-\cosh(a_z\tau))
\label{R_mu1}
\end{equation}

We obtain for the tensor $R^{\mu}v^{\nu}$

\begin{equation}
R^{\mu}v^{\nu}=\left(
\begin{array}{llll}
 -a_z^{-1}\sinh(a_z\tau)\cosh(a_z\tau) & 0 & 0 & -a_z^{-1}\sinh^2(a_z\tau) \\
 d\cosh(a_z\tau)                       & 0 & 0 & d\sinh(a_z\tau)           \\
                                     0 & 0 & 0 &                        0  \\
 a_z^{-1}(1-\cosh(a_z\tau))\cosh(a_z\tau) & 0 & 0 & a_z^{-1}(1-\cosh(a_z\tau))\sinh(a_z\tau) \\
\end{array}\right)
\label{R_mu_v_nu}
\end{equation}

The tensor $R^{\nu}v^{\mu}$ is the transpose of (\protect\ref{R_mu_v_nu}),
so we obtain for $R^{\mu}v^{\nu}-R^{\nu}v^{\mu}$

\begin{equation}
R^{\mu}v^{\nu}-R^{\nu}v^{\mu}=\left(
\begin{array}{llll}
 0                & -d\cosh(a_z\tau) & 0 & a_z^{-1}(1-\cosh(a_z\tau))  \\
 d\cosh(a_z\tau)  &                0 & 0 & d\sinh(a_z\tau)             \\
 0                &                0 & 0 & 0                           \\
 -a_z^{-1}(1-\cosh(a_z\tau)) & -d\sinh(a_z\tau) & 0 & 0                \\
\end{array}\right)
\label{R_mu_v_nu-R_nu_v_mu}
\end{equation}

The scalar $R^{\sigma}v_{\sigma}$ is

\begin{equation}
R^{\sigma}v_{\sigma}=a_z^{-1}\sinh(a_z\tau)
\label{R_sigma_v_sigma}
\end{equation}

Dividing (\protect\ref{R_mu_v_nu-R_nu_v_mu}) by
(\protect\ref{R_sigma_v_sigma}) and taking the derivative with respect
to $\tau$, we obtain

\begin{equation}
\frac{d}{d\tau}\frac{R^{\mu}v^{\nu}-R^{\nu}v^{\mu}}{R^{\sigma}v_{\sigma}}=a_z\left(
\begin{array}{llll}
 0 &\frac{da_z}{\sinh^2(a_z\tau)}&0&\frac{-1}{2\cosh^2(\frac{a_z\tau}{2})}\\
 \frac{-da_z}{\sinh^2(a_z\tau)}&0&0&0                                     \\
 0                &                0 & 0 & 0                              \\
 \frac{1}{2\cosh^2(\frac{a_z\tau}{2})}&0&0&0                              \\
\end{array}\right)
\label{tau_derivative}
\end{equation}

So we obtain the EM field tensor

\begin{equation}
F^{\mu\nu}=\pm\frac{q_1 a_z^2}{\sinh(a_z\tau)}\left(
\begin{array}{llll}
 0 &\frac{da_z}{\sinh^2(a_z\tau)}&0&\frac{-1}{2\cosh^2(\frac{a_z\tau}{2})}\\
 \frac{-da_z}{\sinh^2(a_z\tau)}&0&0&0                                     \\
 0                &                0 & 0 & 0                              \\
 \frac{1}{2\cosh^2(\frac{a_z\tau}{2})}&0&0&0                              \\
\end{array}\right)
\label{em_tensor_exact1}
\end{equation}

and by using the equality $\sinh(x)\equiv 2\cosh(x/2)\sinh(x/2)$ we may
simplify it to

\begin{equation}
F^{\mu\nu}=\mp\frac{q_1 a_z^2}{\sinh(a_z\tau)2\cosh^2(\frac{a_z\tau}{2})}\left(
\begin{array}{llll}
 0 &\frac{-da_z}{2\sinh^2(\frac{a_z\tau}{2})}&0&1  \\
 \frac{da_z}{2\sinh^2(\frac{a_z\tau}{2})}&0&0&0    \\
 0    &     0 & 0 & 0                              \\
 -1   &     0&  0 & 0                              \\
\end{array}\right)
\label{em_tensor_exact2}
\end{equation}

Now we have to find $\tau$ which satisfies (\protect\ref{ret_adv_condition})

\begin{equation}
R^{\sigma}R_{\sigma}=d^2+2a_z^{-2}(1-\cosh(a_z\tau))=0
\label{ret_adv_condition1}
\end{equation}

By using the equality $\cosh(x)-1\equiv 2\sinh^2(x/2)$ we find that
$\sinh^2(a_z\tau/2)=(da_z)^2/4$, and $\tau$ is found by solving
(\protect\ref{ret_adv_condition1}) for $\sinh(a_z\tau)$

\begin{equation}
\sinh(a_z\tau_{\tt RET/ADV})=\mp da_z\sqrt{1+\left(\frac{da_z}{2}\right)^2}
\label{tau_ret_adv}
\end{equation}

We remark that both retarded and advanced fields are identical for
our case, because by setting (\protect\ref{tau_ret_adv}) in 
(\protect\ref{em_tensor_exact2}), the signs cancel.

We obtain, from these results, the EM tensor field

\begin{equation}
F^{\mu\nu}=\frac{q_1 a_z}{2d\left[1+\left(\frac{da_z}{2}\right)^2\right]^{3/2}}\left(
\begin{array}{llll}
 0 &-\frac{2}{da_z}& 0 & 1  \\
 \frac{2}{da_z} & 0 &0&0    \\
 0    &     0 & 0 & 0                              \\
 -1   &     0&  0 & 0                              \\
\end{array}\right)
\label{em_tensor_exact3}
\end{equation}

We calculate now the 4-force on the charge $q_2$. The velocity $v$ of
charge $q_2$ is zero in our reference frame, as we defined before, so
the 4-velocity $v^{\mu}$ is $(1,0,0,0)$.
The force on $q_2$ is then

\begin{equation}
f^{\nu}_2=q_2 v_{\mu}F^{\mu\nu}=\frac{-q_1 q_2 a_z}{2d\left[1+\left(\frac{da_z}{2}\right)^2\right]^{3/2}}\left(
\begin{array}{llll}
 0 &-\frac{2}{da_z}& 0 & 1  \\
\end{array}\right)
\label{force_on_q_2}
\end{equation}

Similarly if we calculate the force on the charge $q_1$, caused by $q_2$,
we just have to reverse the $x$ axis, so we obtain

\begin{equation}
f^{\nu}_1=\frac{-q_1 q_2 a_z}{2d\left[1+\left(\frac{da_z}{2}\right)^2\right]^{3/2}}\left(
\begin{array}{llll}
 0 &\frac{2}{da_z}& 0 & 1  \\
\end{array}\right)
\label{force_on_q_1}
\end{equation}

The force on the whole system is therefore

\begin{equation}
f^{\nu}=f^{\nu}_1+f^{\nu}_2=\frac{-q_1 q_2 a_z}{d\left[1+\left(\frac{da_z}{2}\right)^2\right]^{3/2}}\left(
\begin{array}{llll}
 0 & 0 & 0 & 1  \\
\end{array}\right)
\label{force_on_q_1q_2}
\end{equation}

The only component of the force is in the direction of the
acceleration $z$

\begin{equation}
f_z=\frac{-q_1 q_2 a_z}{d\left[1+\left(\frac{da_z}{2}\right)^2\right]^{3/2}}
\label{force_z_exact}
\end{equation}

For an ideal dipole, according to (\protect\ref{ideal_dipole_condition}),
$da_z \rightarrow 0$, and the force is proportional to the acceleration:

\begin{equation}
f_z=\frac{-q_1 q_2 a_z}{d}\equiv Ma_z
\label{force_ideal_dipole}
\end{equation}

We defined here $M\equiv -q_1 q_2/d$, and dealing with $q=q_1=-q_2$,
$M$ comes out $q^2/d$ and is a positive quantity.

We remark that the configuration defined in Figure
\protect\ref{accelerating_dipole} cannot radiate, and we see that
it behaves like a {\it simple} constant mass. That is why we did not
have to bother whether to use retarded or advanced field, and both
fields gave the same result (see remark after (\protect\ref{tau_ret_adv})).
Another way of understanding this is by the fact that
$F_{self}^{\mu}=(2/3)q^2(\dot{a}^{\mu}-a^{\lambda}a_{\lambda}v^{\mu})\equiv 0$
for the world line described in (\protect\ref{Z_mu}).

Now, does this mass represent inertia? Yes, because if some obstacle is put
in the way of the accelerating body, this obstacle will ``feel'' the
force $F=Ma_z$, exactly as it will feel it for any non-electrical
body of mass M. The obstacle will be hit by an accelerating body,
and will absorb the force $a_z q^2/d$, hence the dipole has inertial
properties of magnitude $q^2/d$.

However, the problem as exposed above, describing an accelerating dipole
is not completely defined, because we must say what caused it to
accelerate. But we can look at it from a different point of view:
suppose we observed a {\it static} dipole from a frame accelerating
in the $-z$ direction.

Any mass $M$ that we observe from an accelerating system will appear
to us as being driven by an {\it imaginary} force (we call it imaginary
{\it aposteriori}, knowing we are not in an inertial system, but we observe it
as a real force). In other words, if the dipole would {\it not} have EM
mass, we would {\it not have observed} the force. And again, we can measure
this mass by dividing the imaginary force $F$, in our case $a_z q^2/d$ by the
acceleration $a_z$, and obtain $M=q^2/d$.\\

\noindent{\bf\large 6. DISCUSSION}\\

\noindent
The purpose of this work was to clarify the energy transfer
process between charged bodies and the world surrounding them.

We showed here that the ``self'' retarded field of a charge, acting on its
neighbor, generates for a radiating antenna a ``self'' voltage which, divided
by the current results in the radiation resistance ($R$), and similarly
the ``self'' advanced field of a charge, acting on its neighbor, generates
for an absorbing matched antenna a ``self'' voltage which, divided by the
current results in {\it minus} the radiation resistance ($-R$), i.e.
acts like current source with internal resistance $R$.

We showed also that the radiation process of an antenna consists of two
energy transfer processes: energy transfer from the current source to the
charges (expressed by $-2/3q^2\dot{a}^{\mu}v_{\mu}$), and energy transfer
from the charges to the surrounding space (expressed by
$2/3q^2a^{\mu}a_{\mu}$). Those
two processes are not in phase, and therefore the mass of the antenna
fluctuates at twice the radiation frequency.

Similarly, the absorption process of an antenna consists of two
energy transfer processes: energy transfer from the surrounding space to
the charges, and energy transfer from the charges to the resistive
absorber. In this case the mass of the antenna fluctuates at twice the
absorption frequency, too.

In a transient period between absorption and radiation, we found out
that during absorption the mass of the antenna increases by the amount
of energy absorbed, and during the radiation period the mass of the
antenna decreases by the amount of energy radiated.

Another interesting conclusion is that bound charges, always appearing
in nature as dipoles, are a manifestation of energy (inertia).

In a philosophical way, one may say that charge itself can be considered
as a manifestation of the interaction process. This may be understood in
the following way. From a macroscopic point of view, an antenna is completely
neutral in the absence of interaction. The charges of course exist all
the time, but they annihilate each other. During the interaction, the charges
form dipoles, and appear as mass fluctuations.

\end{spacing}
\end{document}